\documentstyle[prl,aps,multicol,epsf]{revtex}
\begin{document}
\widetext
\draft
\title{Magnetic stress as a driving force of structural distortions: the case of CrN}
\author{Alessio Filippetti and Nicola A. Hill}
\address{Materials Department, University of California,
Santa Barbara, CA 93106-5050}
\maketitle
\begin{abstract}
We show that the observed transition from rocksalt to orthorhombic 
P$_{nma}$ symmetry in CrN can be understood in terms of stress anisotropy.
Using local spin density functional theory, we find that the 
imbalance between stress stored in spin-paired and spin-unpaired Cr nearest neighbors 
causes the rocksalt structure to be unstable against distortions and justifies the 
observed antiferromagnetic ordering. This stress has a purely magnetic origin, and may be 
important in any system where the coupling between spin ordering and structure is strong.
\end{abstract}


\begin{multicols}{2}

\narrowtext
The simultaneous occurence of structural distortion and change of magnetic ordering
is observed in several magnetic compounds. In general, the interplay between structural and 
spin degrees of freedom depends on the detailed configuration of the electronic and phonon 
states, and a unique picture that elucidates the coupling mechanism is missing. 
A large class of systems where this interplay is strong are materials 
containing ions with degenerate (usually $e_g$) orbitals (Jahn-Teller ions), such as  
Mn$^{3+}$, Cr$^{2+}$ or Cu$^{2+}$\cite{ks,hu}. The degeneracy causes instabilities that can be 
relieved by structural distortions or orbital (and eventually charge) ordering.
In this paper we focus on CrN, one of the most widely employed materials in coating technology, 
due to the extreme hardness of CrN films. It is also the prototype material of strong magneto-structural
interactions, since here, more than in any other case, the coupling is clearly revealed: at room 
temperature it is paramagnetic (PM) in the RS structure, like most of the transition metal nitrides, but at 
N\'eel temperature $T_N$=273-286 K, a both structural and magnetic transition to an antiferromagnetic 
(AFM) orthorhombic P$_{nma}$ phase is observed\cite{cor,blsm,sub,ibcy}.

But other compounds presents indications of strong spin and structural coupling.
To name a few of them, in the perovskite LaMnO$_3$ the distortions are necessary to stabilize the observed 
A-type antiferromagnetic ordering. The doped cobalt perovskite La$_{1-x}$Sr$_x$CoO$_3$ exhibits unusual 
properties as a function of doping and temperature\cite{lsco}, and in the diluted magnetic semiconductors 
Zn$_{1-x}$Cr$_x$Se, the magnetic Cr$^{2+}$ impurities are strongly affected by Jahn-Teller distortions\cite{zcs}. 
Also large is the class of materials manifesting magnetoelastic effects, such as changes of 
magnetic anisotropy coupled to distortions and strains. For example, the martensitic phase transition 
of Ni$_2$MnGa from cubic Heusler to tetragonal structure is accompanied by a change in magnetic 
anisotropy and driven by magnetoelastic coupling.\cite{nmg}. In the layered Cu/Co/Ni/Cu/Si(001) the 
magnetic anisotropy can be induced by strain and controlled through the film thickness\cite{layer},
and in CeNiSn the magnetic ordering can be changed by application of uniaxial strain\cite{cns}.

For all of these materials, the stress is an important quantity, being 
naturally able to bridge electronic and structural properties. The stress relief argument is 
very popular in surface physics to explain processes like surface reconstructions, adsorbtions 
and growth: after a structural perturbation, a stress excess is produced in the system 
(e.g. the surface stress), that is relieved by further structural rearrangements 
involving symmetry breaking and eventually change of atomic density at surface.
In this paper we show that is possible to apply similar arguments to magnetic systems. 
Specifically, we use the concept of stress relief to explain the 
stability of one magnetic ordering with respect to another, and to show how the change of 
ordering is related to a structural distortion.

In analogy to the surface stress, we define {\it magnetic stress} to be the 
stress associated with a change of spin ordering. If $O^1$ and $O^2$ represent two magnetic 
phases of a system with the same structure, the magnetic stress $\hat{T}$ associated with the 
transition is:

\begin{eqnarray}
\hat{T}=\hat{T}^{O^1}-\hat{T}^{O^2}=
{\partial E^{O^1}\over \partial \hat{\epsilon}}-
{\partial E^{O^2}\over \partial \hat{\epsilon}},
\label{equa1}
\end{eqnarray}

\noindent
where $\hat{\epsilon}$ is the strain tensor. $\hat{T}$ has a purely magnetic origin, since it
originates from changes in spin orientations, not structural transitions, but may be relieved 
by structural transitions. We also define the magnetic stress per bond by mapping $\hat{T}$ 
onto the Heisenberg model,

\begin{eqnarray}
\hat{T}= - \sum_{ij}S_i\cdot S_j\> \hat{t_{ij}},
\label{equa2}
\end{eqnarray}

\noindent
where $\hat{t_{ij}}$ is the strain derivative of the usual energy exchange 
interaction parameter $J_{ij}$, and represents the change of stress caused by turning the spin 
orientation on atoms $i$ and $j$ from parallel to antiparallel. A positive
$\hat{t_{ij}}$ means that the single spin-flip produces a tensile stress that can be relieved
by shortening the $i$-$j$ distance. A negative $\hat{t_{ij}}$, instead, represents a compressive 
stress that is relieved by stretching the $i$-$j$ distance.
The meaning of the definitions of Eqs.\ref{equa1} and \ref{equa2} will be transparent when
applied to the practical case of CrN. 

\begin{figure}
\epsfxsize=9cm
\centerline{\epsffile{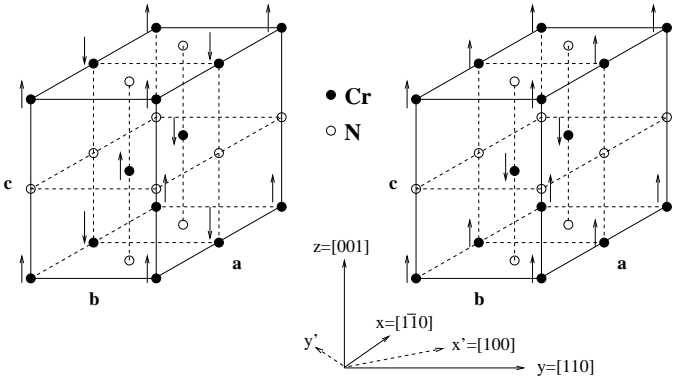}}
\caption{Orthorhombic cell of CrN (4 formula unit) in the observed AFM$^2_{[1\overline{1}0]}$ spin ordering 
(left) and in the AFM$^1_{[1\overline{1}0]}$ ordering (right). 
The cartesian axes of the cubic cell 
(x' and y') are rotated 45$^o$ with respect to x and y.}
\label{fig1}
\end{figure}

Below $T_N$, CrN is orthorhombic, with an unusual magnetic ordering\cite{cor} 
(see Fig.\ref{fig1}, left) built up by (1$\overline{1}$0) FM layers 
whose spin direction alternates up and down after each 2 layers, so that the system 
is antiferromagnetic (AFM) overall (in the following this phase is labeled AFM$^2_{[1\overline{1}0]}$).
Along with the AFM$^2_{[1\overline{1}0]}$, we consider two other AFM phases as possible 
competitors: the AFM$^1_{[1\overline{1}0]}$, similar to the former but with 
a one-by-one compensation of FM layers along [1$\overline{1}$0] (Fig.\ref{fig1}, right), and the 
AFM$^1_{[111]}$, made by (111) FM layers alternated along [111], common for some strongly 
ionic transition-metal oxides (e.g. MnO).

Our calculations are based on local spin density functional theory (LSD)\cite{vxc}, 
and employ plane-wave basis and Troullier-Martins pseudopotentials\cite{fuchs}
(this methodology already described successfully some properties of CrN \cite{ppkb,sss,mmsd,fp}). 
The microscopic stress tensor is calculated in LSD\cite{nm} (within the framework of the ABINIT\cite{abinit} 
project) and problems of finite-size basis have been efficiently circumvented\cite{gonze}.

To rationalize the RS-to-orthorhombic transition, we can picture it as a three step process: 
first, an hydrostatic expansion of the volume accompanies the non spin polarized-to-spin polarized 
transition. This is the dominant step in terms of energy stabilization. 
Second, a further stabilization is obtained from the competition of different magnetic orderings with 
the same (cubic) structure. Third, a shear strain is applied onto the (001) plane, in accordance with 
the experimental finding\cite{cor}. The observed distortion consists of a reduction of the lattice 
parameter $a$ and an increase of $b$ (of $\sim$ 2\%) with respect to their ideal values 
($a=a_0\sqrt{2}$, $b=b_0/\sqrt{2}$, and $c=a_0$, where $a_0$ is the cubic lattice constant), in a 
way that the Cr-N distance is unchanged. Although a second-order effect with respect to 
the hydrostatic expansion, we will show that the distortion 
is fundamental in stabilizing the observed AFM$^2_{[1\overline{1}0]}$ phase.

\begin{figure}
\epsfxsize=6cm
\centerline{\epsffile{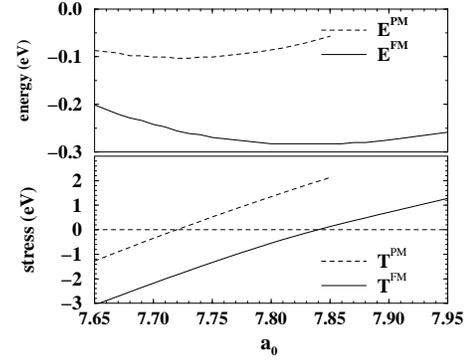}}
\caption{Energies and stresses (per formula unit) of PM and FM phases of RS CrN, as a function 
of the lattice constant $a_0$.} 
\label{fig2}
\end{figure}

The effects related to the first step are described in Fig.\ref{fig2}, where the energies and stresses
of PM and FM orderings in the RS structure are shown. In the PM phase, the high density of states 
(DOS) at the Fermi energy (E$_F$) strongly drives CrN towards a Stoner instability. 
Although the high DOS peak at E$_F$ disappears in the FM phase, the spin splitting does not manage to 
open a gap, thus, 
FM CrN is still a (poor) metal\cite{fp}. It is also a robust magnetic compound, with a magnetic moment 
(M=2.09 $\mu_B$ per Cr atom) almost entirely due to the Cr $t_{2g}$ splitting, and a chemical configuration 
close to Cr$^{2+}$N$^{2-}$.

We find a difference of 0.182 eV/formula unit between PM and FM 
energies, calculated at the equilibrium $a_0$ of the respective phases, that are 7.72 a.u. and 7.84 a.u. 
for PM and FM, respectively. Thus, after polarization, the lattice constant increases $\sim$ 1.5\%.
This strong magnetovolume effect is related to a large magnetic stress. At the PM lattice
constant (where T$^{PM}$ is zero by definition), the spin-polarization produces a compressive 
stress T$^{FM}$=1.89 eV/formula unit, whose relief corresponds to an energy gain 
$\Delta E^{FM}$=T$^{FM}$\,$\Delta$ a/a$\simeq$ 0.03 eV/formula unit.

The increase in compressive stress is a consequence of some charge localization 
(mostly Cr $d$ $t_{2g}$ charge) around the Cr nucleus, due to the spin polarization.
The charge localization increases the kinetic stress, and, in turn, decreases 
the tensile electron-ion interaction.

Within this expanded volume, but still retaining the RS symmetry, we consider 
the competition among different magnetic orderings (the calculated lattice constant is quite 
similar for FM and AFM, i.e. a change of magnetic ordering does not produce that significant hydrostatic 
compression). We find (Table \ref{tavola1}) all the AFM phases are favored over the FM one. 
Moreover, the AFM$^1_{[1\overline{1}0]}$, and 
not the observed AFM$^2_{[1\overline{1}0]}$, is the stable phase in the RS structure. 

\begin{figure}
\epsfxsize=7cm
\centerline{\epsffile{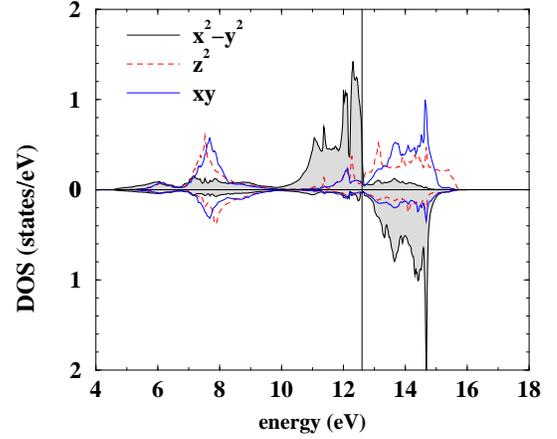}}
\caption{Orbital-resolved DOS of one Cr ion in the AFM$^2_{[1\overline{1}0]}$ CrN (undistorted). 
Due to the 45$^o$ rotation of the (001) plane, $d_{xy}$=$d_{{x'}^2-{y'}^2}$ and 
$d_{x'y'}$=$d_{{x}^2-{y}^2}$, where x' and y' are the axes of the cubic cell (see Fig. \ref{fig1}).
The vertical line corresponds to the Fermi energy.} 
\label{fig3}
\end{figure}

The density of states (DOS) of a single Cr ion in the AFM$^2_{[1\overline{1}0]}$ phase is shown in 
Fig. \ref{fig3}. The Cr configuration is nearly Cr$^{2+}$, with magnetization M=2.12 $\mu_B$.
Notice that in the (001) plane of the orthorhombic cell, the $t_{2g}$ orbitals (whose DOS is drawn by
the shaded area in the Figure), not the $e_g$s, are directed along $\hat{x}$ and $\hat{y}$.  
The $e_g$ orbitals are non degenerate, due to the AFM symmetry, and each of them is occupied by nearly 
half an electron. Their total contribution to the magnetization is only $\sim$ 0.2 $\mu_B$. The exchange
splitting for $t_{2g}$ orbitals is $\sim$ 2 eV, with the $t_{2g}^{\uparrow}$ almost completely filled. 
The global $t_{2g}$ occupation is 2.76, and the contribution to the magnetic moment is 1.9 $\mu_B$. 

We can understand the competition among spin orderings by a two-parameter 
Heisenberg model, taking into account the nearest neighbor Cr-Cr ($J_1$), and the next nearest neighbor
Cr-N-Cr interactions ($J_2$). The calculated parameters are $J_1$=-9.5 meV, and $J_2$=4 meV.
For the Cr-N-Cr interaction there is competition between superexchange (AFM) and double-exchange (FM)
mechanisms involving Cr $e_g$ and N $p$ orbitals ($dp\sigma$ hybridization). The resulting coupling
is slightly FM (i.e. $J_2$ is positive). 
The direct Cr-Cr coupling, instead, is mediated by $t_{2g}$ orbital interactions ($dd\sigma$), 
and is AFM (i.e. $J_1$ is negative), as we may argue assuming, as usual, the Hund's rule.
The formal equivalence of Cr$^{2+}$ and Mn$^{3+}$ ions may suggest some similarities between 
CrN and LaMnO$_3$. Indeed, in both of them the magnetic ion is surrounded by an octhaedral cage
of cations. However, in LaMnO$_3$ the Mn coupling via oxygens is the only relevant interaction,
whereas, for CrN the Cr-N-Cr coupling is smaller than the direct Cr-Cr AFM coupling. 
As a consequence, the stability increases with the number of AFM nearest neighbors 
(8 for the AFM$^1_{[1\overline{1}0]}$, 6 for AFM$^2_{[1\overline{1}0]}$ and AFM$_{[111]}$). 
Also, due to the symmetry of the Cr fcc sublattice, no other arrangement with 
more than 8 AFM nearest neighbors is possible. In other words, the signs of $J_1$ and $J_2$ rule out 
the possibility that other orderings, not taken into account here, might be more stable.

On top of the RS structure, we finally apply a shear strain $\epsilon$ onto the (001) 
plane of the competing phases AFM$^1_{[1\overline{1}0]}$ and  AFM$^2_{[1\overline{1}0]}$
(Table \ref{tavola2}). First consider the stress values for the undistorted (i.e. $\epsilon=0$) 
structure: the AFM$^2_{[1\overline{1}0]}$ phase is under a condition of appreciable stress 
with respect to the AFM$^1_{[1\overline{1}0]}$. This stress originates from
the spin ordering asymmetry on the (001) plane: in the AFM$^2_{[1\overline{1}0]}$ phase, 
each Cr has two spin-antiparallel nearest neighbors along $\hat{x}$, and two spin-parallel 
along $\hat{y}$. The bonds between AFM neighbors are under condition of tensile stress 
(i.e. $T_{xx}>0$), whereas the bonds between FM neighbors store compressive stress
(i.e. $T_{yy}<0$). Clearly, a planar distortion will be able to relieve the stress and lower the energy 
of the AFM$^2_{[1\overline{1}0]}$.
In contrast, the AFM$^1_{[1\overline{1}0]}$, which is isotropic on (001), cannot undergo the same 
stabilization: each Cr has two spin-parallel nearest neighbors in both $\hat{x}$ and $\hat{y}$ 
directions, and the eventual energy gain due to the stretch along $\hat{y}$ would be lost due 
to the compression along $\hat{x}$. 
Also, $T_{zz}$ is different for the two phases: in the AFM$^2_{[1\overline{1}0]}$, the larger number 
of FM Cr-Cr bonds having an orthogonal component causes an excess of compressive stress along $\hat{z}$. 

Applying Eq.\ref{equa2} we can express the stress (at $\epsilon$=0) in terms of exchange-interaction 
parameters. For a planar strain on (001), we have three independent parameters  
corresponding to the in-plane Cr-Cr bonds ($t_1$), and the planar and orthogonal components of the 
out-of-plane Cr-Cr bonds ($t_{2x}$ and $t_{2z}$). Our results gives $t_1$=1.04 eV, $t_{2x}$= 0.24 eV,
and $t_{2z}$=0.20 eV. Thus, the parallel-to-antiparallel spin flip causes a significantly large tensile 
stress for the Cr-Cr bonds parallel to (001).

On the basis of the magnetic stress results at $\epsilon$=0 we are able not only to predict 
the occurence of the distortion for AFM$^2_{[1\overline{1}0]}$, but also to estimate the energy gain 
due to the stress relief. Indeed, a 2\% distortion gives an energy gain
$\Delta E = \left(T_{xx}-T_{yy}\right)\,{\delta a\over a}\sim \,0.04\> eV$
which is enough for the AFM$^2_{[1\overline{1}0]}$ to be stabilized over the
AFM$^1_{[1\overline{1}0]}$.
This estimation can be compared with the direct calculation
of energies and stresses at $\epsilon\neq 0$. As expected, the distortion progressively
relieves the stress and stabilizes the observed phase. Our results gives a theoretical distortion 
slightly larger than the experimental value: at $\epsilon=4\%$ most of the planar stress is 
relieved. In contrast, the distortion on the AFM$^1_{[1\overline{1}0]}$ increases the
stress and is energetically unfavorable.

The stabilization of AFM$^2_{[1\overline{1}0]}$ ordering is ultimately due to different 
stress stored in the bonds between spin-antiparallel and spin-parallel Cr nearest neighbors.
Stress asymmetry and ordering of $t_{2g}$ orbitals are intimately related:
since the Cr-Cr interaction occurs via direct $t_{2g}$ coupling and is AFM,
$J_1$ should vanish in the limit of large nearest neighbor distance, whereas a shortening of 
the distance should increase the overlap between $t_{2g}$ orbitals, making $J_1$ even more negative.

\begin{figure}
\epsfxsize=6cm
\centerline{\epsffile{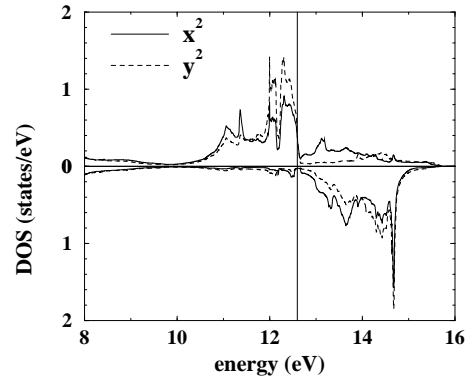}}
\caption{DOS of Cr $d_{x^2}$ and $d_{y^2}$ orbitals in the AFM$^2_{[1\overline{1}0]}$ RS phase. 
The vertical solid line corresponds to the Fermi energy.} 
\label{fig4}
\end{figure}

This explains why a bond contraction stabilizes the spin-antiparallel interactions, while  
a bond stretching stabilizes the spin-parallel interactions. In Fig.\ref{fig4} we show the DOS of 
$d_{x^2}$ and $d_{y^2}$ orbitals (that contribute most to the Cr-Cr planar coupling) 
for the undistorted AFM$^2_{[1\overline{1}0]}$ ordering (obviously they are degenerate for
AFM$^1_{[1\overline{1}0]}$). The tendency of these orbitals to increase
the overlap (and, thus, the hybridization) along $\hat{x}$, and to stay more localized along $\hat{y}$
is clearly visible.

In conclusion, in this paper we have given a definition of magnetic stress, and proposed it as 
a tool for understanding and predicting the subtle relations between magnetic ordering 
and structural properties. For CrN, we show that the coupling between the AFM$^2_{[1\overline{1}0]}$ 
ordering and the orthorhombic $P_{nma}$ structure is driven by the relief of tensile stress stored in the 
AFM Cr nearest neighbors, and associated with the direct interaction of $t_{2g}$ orbitals.
Our results for CrN may stimulate the application of arguments based on stress relief to many other 
magnetic compounds. Indeed, the predictive power of the stress should be important in every case where 
magnetism and structural properties are strongly related.

We acknowledge financial support from the National Science Foundation POWRE program under 
grant number DMR 9973859. Many thanks to Xavier Gonze and the ABINIT group.
 


\begin{table}
\centering\begin{tabular}{ccccc}
          &  AFM$^1_{[1\overline{1}0]}$ & AFM$^2_{[1\overline{1}0]}$ & AFM$_{[111]}$ & FM  \\
\hline
    E     &  0                          &  0.034                     &  0.050        & 0.075   \\
\end{tabular}
\caption{Energies E (in eV/formula unit) of RS CrN for different magnetic orderings,
calculated at the respective theoretical lattice constants.
All the values are relative to that of the AFM$^1_{[1\overline{1}0]}$, that is the most 
stable phase overall, as long as the distortion from cubic symmetry is not considered.
\label{tavola1}}
\end{table}

\begin{table}
\centering\begin{tabular}{c|ccc|ccc}
          & \multicolumn{3}{c}  {AFM$^1_{[1\overline{1}0]}$} &  
\multicolumn{3}{c} {AFM$^2_{[1\overline{1}0]}$} \\
\hline\hline
$\epsilon$  & 0\%  & 2\%    & 4\%            & 0\%    & 2\%    & 4\% \\ 
\hline
    E     &  0    &  0.007 & 0.040           & 0.034    & --0.020 & --0.028  \\
 $T_{xx}$ &  0    &  --0.45& --0.81          & 1.13   & 0.79   & 0.19      \\
 $T_{yy}$ &  0    &  0.41  & 1.03            & --0.95 & --0.55 & --0.27     \\
 $T_{zz}$ &  0    & --0.02 & 0.00            & --0.78 & --0.76 & --0.68     \\
\end{tabular}
\caption{Energies E (in eV/formula unit) and stresses $\hat{T}$ (eV/formula unit)
for AFM$^1_{[1\overline{1}0]}$ and  AFM$^2_{[1\overline{1}0]}$ CrN as a function of
the applied (001) planar strain $\epsilon$. The corresponding changes of lattice
parameters are $\delta a$ = -$\epsilon\, a$, and $\delta b$ = $\epsilon\,b$, i.e. the
$\hat{x}$ axis is compressed, and the $\hat{y}$ is stretched.
All values are relative to that of the undistorted AFM$^1_{[1\overline{1}0]}$ phase.
\label{tavola2}}
\end{table}


\end{multicols}
\end{document}